\documentclass[preprint,12pt]{elsarticle}

\usepackage{amssymb}
\usepackage{amsmath}
\usepackage{float,subcaption}
 \usepackage[numbers]{natbib}

\journal{Mathematical Biosciences}

\begin{document}

\begin{frontmatter}



\title{The Lost Art of Mathematical Modelling}


\author[inst1]{Linnéa Gyllingberg}

\affiliation[inst1]{organization={Department of Mathematics},
            addressline={Box 480}, 
            city={Uppsala},
            postcode={SE-751 06}, 
            country={Sweden}}

\author[inst2]{Abeba Birhane}

\affiliation[inst2]{organization={Mozilla Foundation},
            addressline={2 Harrison Street, Suite 175}, 
            city={San Francisco},
            postcode={CA 94105}, 
            country={USA}}

\author[inst3]{David J. T. Sumpter}

\affiliation[inst3]{organization={Department of Information Technology, Uppsala University},
            addressline={Box 337}, 
            city={Uppsala},
            postcode={SE-751 05}, 
            country={Sweden}}

\begin{abstract}
We provide a critique of mathematical biology in light of rapid developments in modern machine learning. We argue that out of the three modelling activities --- (1) formulating models; (2) analysing models; and (3) fitting or comparing models to data --- inherent to mathematical biology, researchers currently focus too much on activity (2) at the cost of (1). This trend, we propose, can be reversed by realising that any given biological phenomenon can be modelled in an infinite number of different ways, through the adoption of an pluralistic approach, where we view a system from multiple, different points of view. We explain this pluralistic approach using fish locomotion as a case study and illustrate some of the pitfalls --- universalism, creating models of models, etc. --- that hinder mathematical biology. We then ask how we might rediscover a lost art: that of creative mathematical modelling. 

This article is dedicated to the memory of Edmund Crampin.

\end{abstract}



\begin{keyword}
mathematical biology \sep hybrid models \sep critical complexity \sep machine learning \sep equation-free approaches
\end{keyword}

\end{frontmatter}


\section{Introduction}
\label{sec:intro}

The challenges in mathematical biology can be roughly broken down in to three activities: (1) formulating models; (2) analysing models; and (3) fitting or comparing models to data. These activities are part of a larger modelling cycle --- where modellers work together with biologists to try to better understand the study system --- but within that cycle, most of the time, the modeller will be found conducting one of these three activities. Research in mathematical biology has evolved a great deal over the last decades, in particular in response to the rise of machine learning (ML). Indeed, the ML approach --- with its emphasis very clearly on activity (3), that of predicting future data --- can be seen as a challenge to the essence of the research area. We need to find ways of reconciling a mathematical biology approach, largely built on describing biological mechanisms, with rapid progress in predicting patterns in data \cite{baker2018mechanistic}.

We argue that in response to the rise of ML, mathematical biology needs to refocus on activity (1), the formulation of new models. We start, in the next section, by defining an inherent feature of biological systems, that they are complex. Our definition of complexity differs from (is more radical than) those most often provided by modellers, in that it emphasises the open-ended nature of biological systems. In section~\ref{allur}, we critique one approach to complex systems, that of unification. This leads us, in section \ref{sec:open}, to propose another approach to modelling biological systems; one which emphasises a plurality of models.

We then argue, in section 5, that (whether researchers are aware of it or not) the unification and pluralistic approaches emphasise different values. Unification emphasises activity (2), that of analysing models, while plurality emphasises activity (1). We argue that currently, the universalist approach dominates and creation of new models, which is inherent to pluralism, is not sufficiently emphasised. This brings us to, in sections 6 and 7, a discussion of how mathematical biology has responded with the rise of machine learning. We argue that ML, which emphasises prediction (activity 3), is ill-prepared to deal with complexity without incorporating some form of mechanistic model building. But we also, more controversially for those working in mathematical biology, emphasise how some of the responses to the rise of ML have fallen in to the trap of making models of models (or fitting models to data generated by models) rather than innovating by creating new models of biology itself.

We conclude that mathematical biology needs less unification and less analysis of existing models, and more creativity and more creation of new models. We should be creative without fear of them being wrong or producing ideas that are mathematically intractable, with an aim of providing a multitude of tools for better understanding of biological systems.

\section{What is complexity?}
\label{sec:whatiscomplex}

Biological systems are complex systems. This statement is so often made, that it can obscure just how radical the consequences of complexity are for the life sciences. To explain why we say radical, consider one of the most common uses of the term complex systems. In physics and applied mathematics, complex systems science has become a name for a set of modelling tools: networks, power laws, phase transitions and the like which purport to capture general properties of systems. This is explicitly {\it not} what we mean by complexity. Although complex systems models will come up in this article, we do not consider them useful in defining complexity itself.

\begin{figure}[t] \centering 
\includegraphics[scale=0.75]{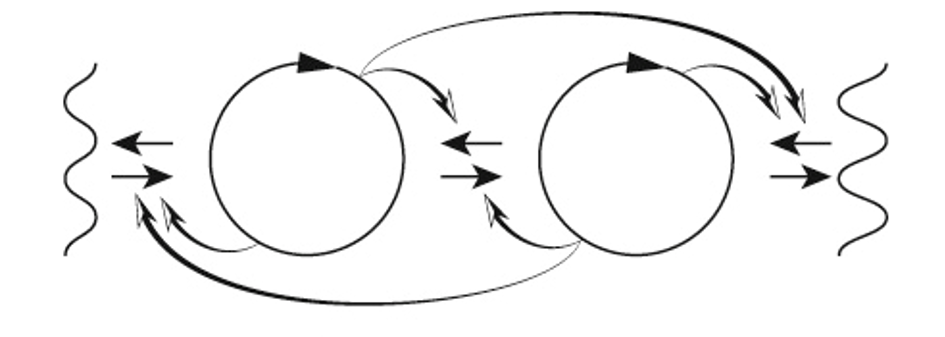}
\caption{The agents (circles) in a complex biological system interact (straight arrows) with each other, their environment (wavy lines), which is partially open and ever-changing, and these interactions are continually adjusted (curved arrows) by the agents themselves. Adapted from Di Paolo et al. (2018) \cite{di2018linguistic}.}
\label{whatiscomplex}
\end{figure}

Instead, the radical definition of complex systems comes from, what is known as, critical complexity. Work by Paul Cilliers and Alicia Juarrero warned against aggrandising models (even supposedly complex systems models) \cite{cilliers2002complexity,juarrero2000dynamics}. They emphasise the need to embrace the ambiguous, messy, fluid, non-determinable, contextual, and historical nature of complex systems. They describe complex phenomena as unfinalizible and inexhaustible, which means that  we can never capture any given biological system entirety with models \cite{cilliers2016we}. Figure \ref{whatiscomplex}, adapted from Di Paolo et al. (2018), captures the interdependence, fluidly and interactivity of agents and environments in a complex system \cite{di2018linguistic}.
Complex systems are open-ended, which means there is no uncontested way of telling whether what we have included in a model is crucial or what we have omitted as irrelevant is indeed so. Models can, according to the critical complexity approach, be contradictory: we can accept two incompatible predictions as both describing the same system. 

This approach views a model as a snapshot of a system and no single snapshot tells the whole story. For modelling the human body, for example, ``a portrait of a person, a store mannequin, and a pig can all be models” \cite{blanchard2012differential}. None is a perfect representation, but each can be the best model for a human, depending on whether one wants to remember an old friend, to buy clothes, or to study anatomy. The critical complexity view suggests that theoreticians should avoid specialising in any one modelling approach and try to find the right set of models to understand a particular system in a given context. 

There can, of course, be more than one definition of complex systems. Indeed, Cilliers and Juarrero's approach to complexity encourages a plurality of definitions (after all, there is no single view of a system). We would, though, emphasise that it is the radical definition of complexity --- in which systems always resist a complete description, are open and unfinalisable --- which is least well understood by mathematical biologists today. It is therefore important to investigate how complexity should be approached in the study of biological systems.

\section{The allure of unification}
\label{allur}

Precisely because most biological systems are more complex than physical systems, they are also more difficult to model. In another article in this collection, Vittadello and Stumpf outline two broad approaches that might be adopted \cite{vittadello2022open}. The first of these approaches builds on the motto put forward by Philip Anderson, that 'More is Different' \cite{anderson1972more}: suggesting that each level of biological organisation requires different types of approaches. The second of these, which we will critique in this section, suggests that the way forward is greater unification and increased mathematical rigour.

In presenting the second approach, that of unification, Vittadello and Stumpf argue that the success of mathematical modelling of physical systems suggest that further progress in mathematical biology can be best made with even more advanced mathematics \cite{vittadello2022open}. With the complexity of biological systems comes a need for rigorous definitions of biological concepts, and they propose a definition-theorem-proof style as a way forward. Accompanying this idea, comes a focus on unification. In the same way as there are unifying theories in physics --- relating to energy conservation, entropy, etc. --- there ought to be unifying models for biological systems. Under this view, an increase in rigour is supposed to tame the complexity of biology. Vitadelli and Stumpf suggest that unification and rigour could lead to avoidance of excessive incrementalism in model development, as well as avoidance of a focus on development of simple models of simple systems. 

The idea of unification in biology has been echoed by many others \cite{rashevsky1969outline, kitcher1981explanatory, borovik2021mathematician,van2007biology}. For example, van Hemmen claims that some of the universal laws of biology might have already been discovered in neurobiology  \cite{van2007biology}. Since mathematical models can describe the behaviour of  biological systems at certain scales, the equations of the models, van Hemmen argues, could be seen as `universal laws'. The question is how to find the appropriate scale for these universal laws. We should, van Hemmen argues, be patient: it has taken humanity hundreds of years to discover the physical laws of the Universe formulated through mathematics; with enough time we will discover the universal laws of biology too \cite{van2007biology}.

Finding the appropriate scales and determining unifying laws is certainly part of modelling biological systems. For example, a fundamental difference between most biological and physical systems is the conservation of momentum. For self-propelled particle models --- which are used for modelling biology on scales ranging from cells to flocks --- momentum is not conserved \cite{czirok2000collective}, and thus classical kinetic theory used in physics to derive macroscopic equations is not applicable. To get around this problem, Degond introduced the Generalised Collision Invariant \cite{degond2008continuum}, from which it is possible to derive macroscopic equations for many self-propelled particle systems used in modelling flocks and other systems in biology \cite{degond2010diffusion,degond2013macroscopic,degond2017alignment,degond2020nematic}. This method allows us to show convergence between self-propelled particle systems on microscopic and macroscopic scale, i.e. between the movement of e.g. flocks of birds described by local interactions between a few individuals, and PDEs describing the angular and velocity distributions of the flock as a group. 

In the example above, an approach like the Generalised Collision Invariant, when applied to the collective motion of real biological systems, usually fails to provide the answers biologists are looking for. Indeed, the very point of self-propelled particle models is to capture the rich, varying dynamics of different schools, swarms and flocks. Derivations of continuum equations for infinitely large populations provide little insight into these questions.  While all of these systems may well share a common invariant, this is not the key issue at hand. Biologists want to find the details of shapes of schooling \cite{herbert2011inferring}, understand how a wave of fish escape from a predator \cite{herbert2015initiation}, find the mechanism behind the V-shape of migrating birds \cite{portugal2014upwash}, measure the sociability of fish schooling based on their movement \cite{sumpter2018using}, study leadership in flocks of pigeons \cite{nagy2010hierarchical, nagy2013context}, or understand the mechanisms behind shepherding sheep \cite{strombom2014solving}; to give just a few examples. We have illustrated this point with a specific example, but see the point as applying more widely. While unification in mathematical biology is tempting, it often neglects the complex nature of biological systems. 

Our discussion of the Generalised Collision Invariant is meant to give one concrete example of how a mathematically appealing universal idea fails to give insight into range of complexity seen in biology \footnote{In this article we take our examples from fish locomotion. We choose an area we understand well in order to illustrate our points in a concrete way. We encourage the reader to imagine similar examples in their own specialised research area.}. Such approaches may well lead to new, interesting and beautiful mathematics \cite{reed2015mathematical}, but there is no reason (a-priori) that they will give deeper biological insight. In biology, experimental results are noisy, non-stationary and often differ across species and scales. Studies of self-propelled particles span species from spiders through fish to humans, as well as sperm and cell interactions. When formalising such models, we have to ask the question, exactly which biological entity or species is it that is being formalised? Is the relevant scale the molecules, the cells, the organs, the animal or the collective? These questions are not amenable to a universal approach or reducible to a small number of equations.

A focus on rigour in biology is an example of over-mathematization, a phenomena frequently discussed in economics \cite{beed1991critique, bouleau2011excessive,bouleau2013can, krugman2009did,  moosa2021mathematization}. The critique has been summarized by the Nobel prize laureate Paul Krugman, who described that ``the economics profession went astray because economists, as a group, mistook beauty, clad in impressive-looking mathematics, for truth." \cite{krugman2009did}. A similar phenomena has occurred in theoretical physics, where the focus on developing beautiful mathematical theories has taken precedent over genuine insight into physics \cite{hossenfelder2018lost}. There is a danger, that in trying to find unification, mathematical biology gets stuck at analysing/unifying simple models, none of which are appropriate for any specific system. Moreover, in search for unification and general methods in biology, we might neglect to study actual systems because they are too complicated or detailed. 

The allure of unification often centres the idea of deriving properties of the collective from interactions between individuals. Countering the possibility of unifying biology through such an approach, Sandra Mitchell argues that microscopic phenomena (cells, molecules, atoms) are not always suited for capturing the rich variety of relations found in biological sciences \cite{mitchell2003biological}. Scientific representations are abstractions or idealizations, and thus only represent partial features of individuals or a system \cite{mitchell2002integrative}. As such,  the abstractions/idealizations do not constitute identical representations across the two levels. Thus, even if the descriptions at each level is accurate, they may, by being partial, not represent the same features of nature. As a result, there is no straightforward derivability or intertranslability relationship between levels \cite{mitchell2003biological}.

Unification is a reductionist approach \cite{mitchell2006integration}. Multilevel, multi-component, complex systems that populate the domain of biology cannot be reduced to a simple, unified picture of scientific theorizing \cite{mitchell2003biological}. And even though contributions to mathematical biology can be made by unification approaches, they do not account for all the explanations that biologists seek \cite{mitchell2006integration}. This view echoes that of Dupré, who explains unification as another way of arguing for the (flawed) reductionist hypothesis that all of science can be reduced to a description based on simple building blocks \cite{dupre1983disunity}. Reductionism fails to account for what Noble calls the `relativity principle', that there is no ``privileged scale at which biological functions are determined'' \cite{noble2012theory}. 

In summary, there are both philosophical (universalism is another form of flawed reductionism) and practical (supposedly general equations don't capture the type of questions biologists ask) arguments against a unification approach to biology. The question now is what is the alternative?

\section{The pluralistic approach}

\label{sec:open}

\begin{figure}[t] \centering 
\includegraphics[scale=0.5]{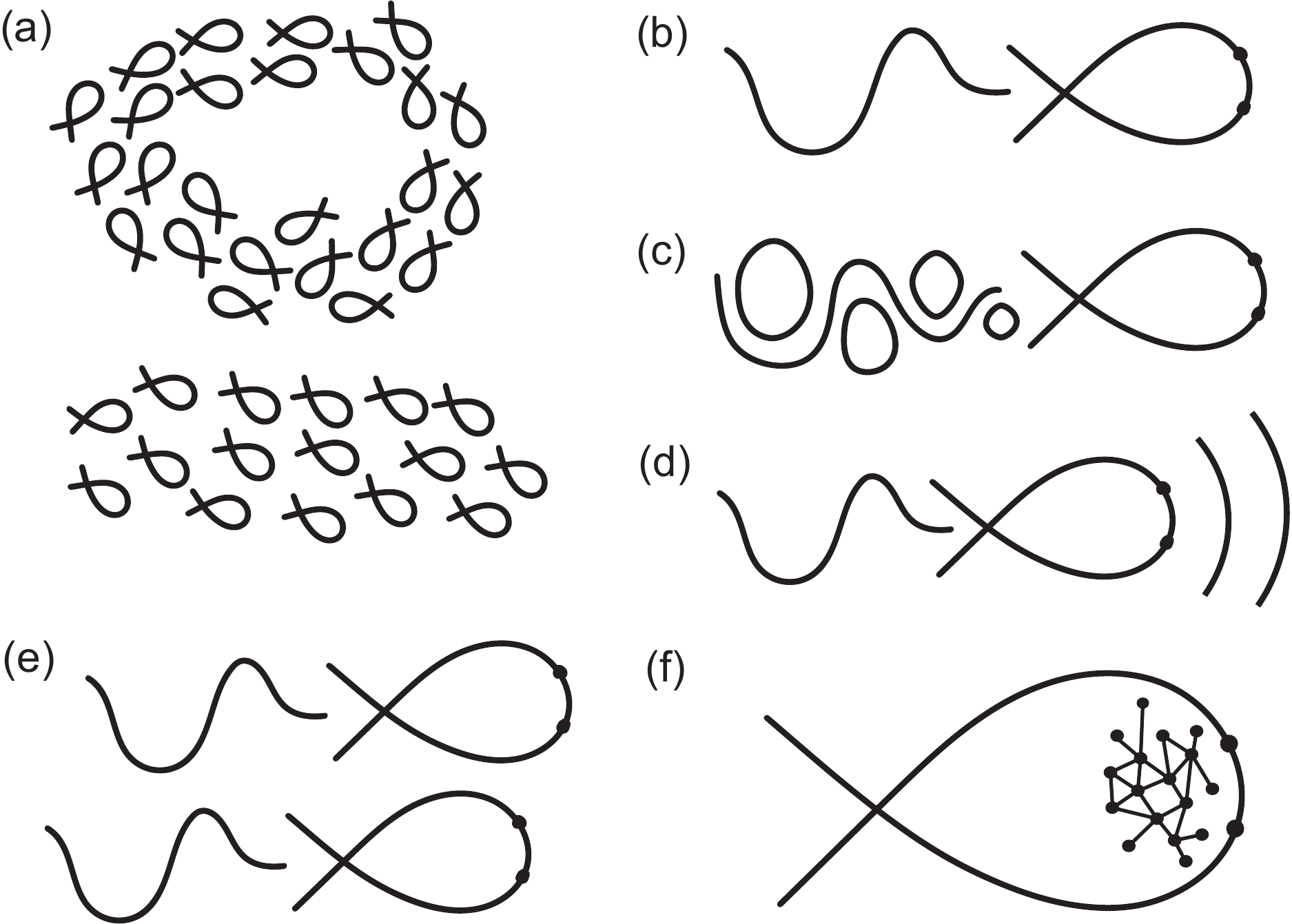}
\caption{Under the Pluralistic Approach to Biology, fish locomotion can be understood and modelled in many different ways. When modelling the collective motion of fish schools, self-propelled particle models can be used to capture and explain their motion (a). However, most fish use burst and glide movement, where they alternate between accelerated motion and powerless gliding (b). This type of motion can be understood from a hydrodynamical perspective (c), or a behavioural ecology perspective (d), as there are both energetic (c) and perceptual benefits (d) of this type of motion. When swimming in pairs, the fish influence each others burst and glide motion (e). Intermittent locomotion can also be understood and modelled through neurobiological perspective, where the activity in the brain affects the motion (f). }
\label{manyfishways}
\end{figure}

Given that biological knowledge is fragmented and that biological systems are complex, we have argued that it is not useful to build a general theoretical framework or to strive for unification. We now outline, in this section, an alternative: the pluralistic approach to modelling biological systems. Pluralism embraces complexity by never seeking to close a biological system in a single or a small number of formalisms, but endlessly endeavours to find new ways of looking at the world. Pluralism reflects the open, unfinalisable nature of complex systems described in section 2.

We illustrate the pluralistic approach by focusing on one specific area of biology, modelling fish locomotion, before broadening out (in the next sections) to explore the lessons we can learn from this approach when looking at other systems. Figure \ref{manyfishways} illustrates various ways that fish locomotion can be modelled. When modelling the behaviour of fish shoals and schools, self-propelled particle models are widely used \cite{partridge1982structure, aoki1982simulation, reynolds1987flocks, huth1992simulation, vicsek1995novel, couzin2002collective, strombom2011collective} (figure \ref{manyfishways}a). The most prominent model, called the Vicsek model \cite{vicsek1995novel}, assumes that each fish (or particle) moves with constant speed, while its direction is updated at each time step to be closer to the average direction of individuals within its neighbourhood. A noise term is added to model uncertainty or error in the fish's direction. These type of models have proven useful when explaining many aspects of schooling behaviour, for example, how body size influences shoaling patterns \cite{romenskyy2017body}, how large schools switch between different organisational states \cite{tunstrom2013collective} and the spread of escape waves in response to a predator \cite{herbert2015initiation}. 

Although self-propelled particles have been central to the study of collective motion, both in fish and other species, there are a several, fundamental ways in which they do not capture the locomotion of fish. For example, rather than updating their angle to turn towards other fish, as assumed in many models \cite{aoki1982simulation, reynolds1987flocks, huth1992simulation, couzin2002collective}, between-fish attraction and repulsion is mainly mediated by speed changes \cite{herbert2011inferring}. Even in the absence of interactions, constant speed is not typical swimming behaviour for most fish species. Zebrafish \citep{kalueff2013towards}, koi carps \citep{wu2007kinematics}, guppies \cite{herbert2017predation}, cod \citep{videler1982energetic}, red nose tetra fish \citep{li2021burst} and many other species swim by alternating between accelerated motion and powerless gliding \citep{weihs1974energetic} (figure \ref{manyfishways}b). 

The discrepancy between how fish are modelled collectively and individually can be seen as an example of logical incompatibility between models \cite{mitchell2003biological,cusimano2019integrative}. After all, fish cannot simultaneously swim both at a constant speed as in the Viscek model and according to a varying speed, as in a burst and glide model! Yet the models in figures \ref{manyfishways}a and \ref{manyfishways}b might view the same fish in both of these (contradictory) ways. 

From a unification point of view, logical incompatibility might be seen as an indication that the models are, at least in some aspects, wrong or that we should try to find ways to unify them to make them both approximately correct (possibly on different time scales). Building on our definition of complexity and open systems, however, we would downplay the importance of investigating relationships between models. From the point of view of complexity, since each model is a different snapshot of a system, taken from a different point of view, using a different camera and lens, we should not be surprised to find an element of incompatibility between models. If we accept the idea of complex phenomena as unfinalizible and inexhaustible (as we outlined in section 2), we cannot expect our models to be logically compatible on every level. Moreover, lack of compatibility should not concern us: it is simply a consequence of taking a different view, of using a different lens, which bends the light of observation in a different way.

Even within the context of intermittent burst and glide motion, we can find several useful and correct, yet logically incompatible approaches to fish locomotion. For example, a range of biomechanical models \citep{weihs1974energetic, videler1982energetic, blake1983functional, fish1991burst, drucker1996use, akoz2017unsteady, floryan2017forces, paoletti2014intermittent} have been proposed for fish locomotion (figure \ref{manyfishways}c). These have been used to show that there are energetic advantages of burst and glide behaviour, when compared to constant swimming speed. The energetic cost of swimming is minimized during the glide phase, where the body is rigid and the fish decelerate due to water resistance \citep{weihs1974energetic}. Other models have focused on a behavioural ecology perspective, since intermittent motion is associated with many ecologically relevant behaviours, e.g. foraging, mating, exploration and predator evasion \cite{wilson2010boldness,andersson1981optimal}. There are also perceptual benefits that can arise from the pauses in locomotion, such as the sensory system's capacity to detect relevant stimuli increases, \cite{kramer2001behavioral, benichou2011intermittent} (figure \ref{manyfishways}d). 

Yet another level to the understanding of intermittent locomotion of fish, is to look at why fish use burst and glide when swimming in pairs. Theoretical studies from hydromechanical perspectives show that there are energetic advantages to intermittent motion in this setting \cite{maertens2017optimal, li2019energetics}. However, the hydromechanical models neglect the social aspects of intermittent locomotion when swimming in pairs (figure \ref{manyfishways}e). In fact, burst and glide swimming have been shown pivotal to detect and quantify social interactions between individual fish \citep{herbert2017predation}. Moreover, high burst speed in response to neighbours evolves when subjected to artificial selection  \cite{kotrschal2020rapid}. Also, when studying intermittent locomotion of pairs of fish, leadership can become apparent \cite{nakayama2012initiative}. For example, in pairs of freely exploring eastern mosquitofish, it is possible to categorize the fish into leaders and followers \cite{schaerf2021statistical}.

The locomotion of fish can also be studied in neurobiological settings (figure \ref{manyfishways}f), leading to questions about how different nerve cells affect the initiation of motion seen in fish \cite{wiggin2012episodic, naumann2016whole, chen2018brain,del2020behavioral}. This can also be linked to genetical aspects of fish locomotion. For example, studies on zebrafish show that genetically encoded calcium-indicators provide a direct link between signalling at a cellular level and functional output in the form of swimming behaviour \cite{del2020behavioral,koning2022deep}.

Above we have listed many different approaches to modelling fish locomotion. Depending on what question we want to answer, the model used is different. It is this creation of many different viewpoints which is pluralism. This multiplicity (or plurality) of ways of seeing a system is even stronger than suggested by Philip Anderson's `More is Different' approach \cite{anderson1972more}. Anderson emphasised that ``psychology is not applied biology, nor is biology applied chemistry'': each level of organisation requires completely new approaches. We would go further, arguing that even within a single biological phenomena, at only one organisational level, we need a whole range of different explanations and models.
 We simultaneously engage many different frameworks and views of a system, each designed to answer a different sub-question. We take different snapshots of the system and then use each of them to construct a bigger picture of the system. The more snapshots we include, the more comprehensive the bigger picture.

Our approach follows, what Sandra Mitchell calls, integrative pluralism \cite{mitchell2002integrative,mitchell2003biological,mitchell2006integration}. Like us, she argues that complexity in nature, particularly in biology, has direct implications for our scientific theories, models, and explanations. To quote, ``nature is complex and so, too, should be our representations of it'' \cite{mitchell2003biological}. Mitchell's own examples build on how social insect biologists explained the emergence of division of labour, by focusing on the effects of causes at one level (genetics, single organisms, and colonies), while idealizing away the other potentially relevant factors \cite{cusimano2019integrative}. She argues that scientists do not need unified theories to provide causal explanations. Nor is there a requirement for logical compatibility between explanations. Mitchell's characterisation of biological research as a whole, we believe, translates even to the mathematical biologist's approach to creating formal models. To paraphrase her: ``biology is complex and so, too, should be our mathematical models of it''. 

Comprehensive, integrated understanding of biology doesn't come from one universal model, but rather the synergy of many different, potentially contradictory models. Having many models helps us understand more. 


\section{How modelling approaches shape mathematical biology}

\label{sec:casestudies}

Our rejecting unification and embracing a pluralistic approach, which emphasises the open nature of biological systems, might be viewed as a purely philosophical exercise. It could also be seen as a question of personal taste. After all, practicing mathematical biologists seldom discuss whether they see the world in terms of unification or integrative pluralism, they just get on with their job... don't they? 

In this section we argue that the formalism and unification approach has a strong influence on what is valued in mathematical biology research. To understand this point, let's return to the three activities outlined in the introduction: (1) formulating models; (2) analysing models; and (3) fitting or comparing models to data. Under a unification approach, which emphasises the importance of mathematical formalism, activity (1) is about finding a small number of universal models which explain as many biological phenomena as possible; and activity (2) requires great care in developing a precise formalism to be clear about the universal properties of the model. Under an integrative pluralism approach activity (1) is about producing lots of different models which view a system through different lenses; and activity (2) is important to get right initially, but details are less important, since we are happy to discard the model once it has told us something useful. 

Unification and integrative pluralism thus emphasise very different values and practices. Whether or not these values are acknowledged by researchers, we can look at the type of activities carried out by mathematical biology researchers and see which approach is more predominant. This is what we do now.

We have already described a pluralistic approach to fish locomotion (figure \ref{manyfishways}) and emphasised the success of self-propelled particle models in describing schooling patterns. However, while there are valuable empirical studies of bird flocks and fish schools, where variations of these models are used to understand the details, these are outnumbered by articles reporting on simulations and investigations of theoretical properties of flocking models \cite{chate2008collective,grossmann2014vortex,szabo2009transitions}. A similar pattern is seen in evolutionary game theory, which provides insight into how spatial and genetic structure is important to the evolution of co-operation \cite{axelrod1981evolution,nowak1992evolutionary,hofbauer1998evolutionary}. This approach has produced countless theoretical questions about how co-operation evolved in different (artificial) settings \cite{nowak2006five, szabo2007evolutionary, perc2010coevolutionary, wang2015evolutionary}, which are detached from observations in the natural world. This is not to say that evolutionary game theory is not useful in biology, it is just to point out that testable predictions have been accompanied by a massive sub-literature simply analysing model properties.

Similarly, complex systems tools --- such as networks \cite{watts1998collective, barabasi1999emergence}, power laws \cite{newman2005power}, phase transitions \cite{domb2000phase} etc. --- often purport to capture general properties of systems and suggest that studying these models will give very general insight, in terms of scaling laws or unifying rules. In making claims of universality, modellers sometimes suggest that biology will succumb, like physics, to an understanding based on one or a small number of models. Under the definition of complexity we use here, the more radical definition, such universality is impossible: complex systems are not complex if they can be reduced to a small number of universal rules. 

Focusing on network science, Fox-Keller argues that claims that scale-free networks and power law distributions are universal laws of life are problematic on two counts \cite{fox2007clash}. Firstly, power laws are not as ubiquitous as was originally supposed \cite{broido2019scale}. Secondly, and more importantly to Fox-Keller, the existence of these distributions tells us nothing about the mechanisms that give rise to them \cite{fox2007clash}. Many reported power laws lack either (or both of) statistical and mechanistic support \cite{stumpf2012critical} There are, at least, a dozen distinct ways to derive power laws from theoretical models \cite{sornette2006critical}, making them far from universal. And asking questions about how power laws should be measured has led to better practices for model fitting (and identifying cases in which they don't fit) \cite{clauset2009power}. Again, it is the details that matter in biology. Power laws don't fit everywhere.

We note that it is activity (1), rather than activity (2), which typically provide the biggest steps forward in Science. Specifically, it is the initial model (prisoners dilemma, chaos theory \cite{may1974biological}, Turing's work on morphogenesis \cite{turing1990chemical}, the Fitzhugh-Nagumo model \cite{fitzhugh1961impulses}, Yule's power law model \cite{newman2005power}, Viscek's SPP model \cite{vicsek1995novel}...) which provides the most inspiration and insight into the biological world, rather than analyses of small variations of these initial models. Yet, universalism, with its emphasis on formalism, prizing activity (2) over activity (1), remains a dominant force in shaping what constitutes mathematical biology.

We believe that the emphasis on model analyses, rather than creating new models, is caused by a tendency towards universalism. The result is  mathematical analysis of small variations of existing models (activity 2) at the expense of creating very novel and different models (activity 1). Admittedly, the observations we make above are qualitative. We have not carried out a comprehensive literature review comparing universalism and pluralism approaches, but instead we appeal to the active mathematical biologist to consider their own research field and think about the theory/application balance. We would imagine for most (leaving activity 3 aside for now) the balance is towards 2, rather than 1. 

This imbalance, we believe, is wrong. Small variations of already existing models seldom provide additional insight into biological systems. We would follow Reed, who wrote in in his 2004 Essay {\it Why is mathematical biology so hard} \cite{reed2004mathematical}: "Don’t do mathematical biology to satisfy a desire to find universal structural relationships; you’ll be disappointed. Don’t waste time developing “methods of mathematical biology”, the problems are too diverse for central methods. What’s left is the biology. You should only do mathematical biology if you are deeply interested in the science itself."

\section{Machine learning can't replace modelling}

The distinction between universalism and pluralism has become particularly important with the move away from activities (1) and (2) ––– model building and analysis ––– towards methods broadly described as machine learning (ML), which emphasise activity (3). Initially, machine learning methods were proposed primarily for data collection --- for example, computer vision was proposed to track movements of fish and cells --- but it later became clear that these methods could also be used to pick out patterns in the data. In an early example, Berman and colleagues showed how different types of fruit fly behaviour (grooming with different legs, running, moving of wings etc.) could be categorised without the need for human definitions of these activities \cite{berman2014mapping}. This work has evolved in to a field of computational ethology, which Pereira recently claimed will ``in the near future, make it possible to quantify in near completeness what an animal is doing as it navigates its environment'' \cite{pereira2020quantifying}. 

Such claims place activity (3), that of fitting or comparing models to data, as central to the scientific endeavour. The proponents of this approach \cite{perretti2013model, roesch2021collocation, rackauckas2020universal} sometimes even go as far as to suggest that activity (1), that of creating models, is now redundant. For example, Rackauckas and colleagues claim that traditional mathematical models are only required because of too small training datasets and that models which are not learnt directly from data have an inductive bias, because they use assumptions about the underlying system being modelled \cite{rackauckas2020universal}.

We reject such claims as yet another form of universalism. As Nurse has recently argued, data should be a means to knowledge, not an end in themselves \cite{nurse2021biology}. Nurse emphasises that the hypothesis free approach of collecting data is just the first step when doing biological research. In order to make advancements in biological sciences, new hypotheses and theories need to be formulated. Reed follows a similar line of argument: ``data itself is not understanding. Understanding requires a conceptual framework (that is, a theory) that identifies the fundamental variables and their causative influences on each other'' \cite{reed2004mathematical}. 

In the context of machine learning, Birhane and Sumpter make a distinction between closed systems --- such as games like Chess and Go, some image analysis tasks and short term nowcasting of the weather --- that are entirely defined by the available data, and open systems --- like fish schools and other biological systems --- similar to those we discuss above, which can be viewed in multiple ways \cite{birhane2022games}. The only systems which can modelled by data alone, in the way Rackauckas and colleagues propose, are those which are fully closed. Indeed, data-driven models, are just representations of the data itself, rather than insights in to that data. To take an example given by Nurse, what if Darwin had just fed in the data of size and shapes of finch beaks into a neural network? The deep learning algorithm would have found clusters and patterns and might have predicted the future development of the beaks on different islands, but would never formulated the theory of natural selection. An ML approach, when employed in the absence of activities (1) and (2), can be characterised as a view from nowhere \cite{birhane2021algorithmic}: without context data analysis becomes meaningless to us, the scientist interpreting the results.

Not only are there (as we have already emphasised) many ways of seeing a system, there are also many other reasons for doing modelling, over and above making predictions. For example, models can also be used to guide data collection for future experiments, or to capture qualitative behaviors of overarching interest and lead to new scientific questions being posed \cite{epstein2008model}. While pure machine learning models are focused on prediction, a large part of biological understanding focuses on mechanistic explanations \cite{braillard2015explanation, mekios2015explanation}. There is no consensus definition of mechanisms, but Illari and Williamson offers the following:  ‘a mechanism for a phenomenon consists of entities and activities organized in such a way that they are responsible for the phenomenon’ \cite{illari2012mechanism}. Thus, for a mathematical model to give a mechanistic explanation to a phenomenon, the model cannot merely summarize and describe the data, but rather the model should encode a mechanism generating the observed phenomenon \cite{breidenmoser2015explanation}. 

Breidenmoser and Wolkenhauer make a distinction between mechanistic models, which explain a system by describing underlying biological processes, and phenomenological models, which only “save the phenomena” by fitting a curve to the data \cite{braillard2015explanation, breidenmoser2015explanation}. The key problem with a purely machine learning based approach is that it says little about the processes, i.e. mechanisms, behind the data \cite{smaldino2017models}, and instead focuses on “saving the phenomena", to use Breidenhoser and Wolkenhauer's term.  Phenomenological models, like many `view from nowhere' machine learning models, might be a good start in understanding statistical relationships between variables, and thus a first step towards modelling a phenomena, but these models do not contribute to a deeper understanding  \cite{breidenmoser2015explanation}.

\section{Are hybrid models the answer?}

Moving away from the idea that machine learning can fully replace mathematical modelling, several authors have proposed integration of mechanistic models and machine learning methods in the form of hybrid models \cite{baker2018mechanistic, vittadello2022open}. These come in many forms, from neural ordinary differential equations \cite{bonnaffe2021neural,roesch2021collocation} and biologically informed neural networks (BINNs) \cite{lagergren2020biologically, daneker2022systems}, to symbolic regression \cite{gaucel2014learning, martin2018reverse} and equation learning \cite{martius2016extrapolation, nardini2021learning}. Hybrid models can both have an underlying specified mechanistic model and then use machine learning methods to infer parts of the equations (as in BINNs \cite{lagergren2020biologically}) or aim to find analytical expressions directly from data (as in equation learning \cite{rackauckas2020universal}). The end product is thus a mechanistic model, in form of, e.g., a dynamical system \cite{nardini2021learning}.  

To give a concrete mathematical example, consider a reaction diffusion equation of the following form 
\begin{equation}
    u_t= \left(\mathcal{D}(u)u_x\right)_x + \mathcal{R}(u), \label{RDeq}
\end{equation}
where $\mathcal{D}$ is a function of $u$ describing the diffusion process, and $\mathcal{R}$, also a function of $u$, describing the reaction process. This equation is used in a range of situations in mathematical biology, from pattern formation, to insect dispersal, to spread of epidemics, to tissue growth. Depending on the application, $\mathcal{D}(u)$ and $\mathcal{R}(u)$ take different forms. For example, in the Fisher-KPP equation, which was originally used to describe the spatial spread of a favoured gene,  $\mathcal{D}$ is a constant and $\mathcal{R}(u)=ru(1-u)$ \cite{murray2002mathematical}.
Choosing the correct form of $\mathcal{D}$ and $\mathcal{R}$ is an open question and the focus of many research efforts \cite{lagergren2020biologically}. The traditional approach for solving this task is to choose an appropriate form of $\mathcal{D}(u)$ and $\mathcal{R}(u)$ based on first principles and then try to fit the parameters of the model to the data. 

Another approach, is to not state the form of $\mathcal{D}(u)$ and $\mathcal{R}(u)$ explicitly, but instead learn the functions directly from data. There are different methods for achieving this. For example, sparse regression \cite{brunton2016discovering, rudy2017data} and theory-informed neural networks  
\cite{lagergren2020biologically} are two of them. Also, $\mathcal{D}(u)$ and $\mathcal{R}(u)$ do not need to be explicitly defined, but can instead just be modelled with data \cite{yates2009inherent}. This approach is sometimes referred to as an equation-free approach as parts of the equations do not need to be explicitly formulated \cite{yates2009inherent}. Other applications of so-called equation free approaches can be found in, for example, in ecosystem forecasting \cite{ye2015equation}. 

Hybrid models are sometimes presented as defining a complete modelling cycle: doing all three activities -- (1) formulating a model, (2) analysing a model and (3) fitting the model to data. But even though parts of the equations do not need to be explicitly formulated in this approach and are directly learnt from the data, the underlying model (in our case the reaction diffusion equation in equation \eqref{RDeq}) is already specified. Thus, equation-free approaches like these are useful for model selection, validation, and analysis, which is part of activity (2) and (3), but they do not replace activity (1). 

Even methods that infer analytical expressions directly from data, with no underlying model like the reaction-diffusion example, cannot be used to replace activity (1). Building mathematical models is more than choosing a set of differential equations that describe data. For example, equation learning would never have formulated inclusive fitness theory in the way Hamilton did after careful across species observations \cite{hamilton1964genetical}. Nor would symbolic regression produce the Vicsek equations and the idea of self-propelled particles solely from analysing videos of bird flocks. 

Thus, while approaches combining machine learning and mechanisms are certainly part of the way forward for mathematical biology, we also need to look critically at whether they are a genuine step away from the universalist thinking, which we have criticised in earlier sections of this article. 

One important critique in this direction arises when researchers apply machine learning methods to models, rather than to data from natural systems. For example, parameters and properties of agent-based models can be learnt using machine learning methods \cite{ten2021use,patsatzis2023data}. In such settings, simulations are seen as a way of generating data on which to test methods for fitting models (i.e to improve the way we perform activity 3). The limitation, from the perspective of a complexity approach, is that simulated data from known models does not come from a complex system (as defined in section 2). Models are not in themselves open-ended or unconstrained, in the same way a biological system is. 
Instead, an already well-established view of a system in the form of one model is studied in the context of another model. This approach implicitly avoids the challenge of formulating new models. 

For example, Roesch and colleagues, apply a collocation based method for training neural ODE:s, i.e. ODE:s where the derivative is learnt directly by a neural network \cite{roesch2021collocation}. To demonstrate the applicability to biology, the neural ODE is trained on data generated from the classical Van der Pol oscillator with added Gaussian noise. The authors view the model as a promising hybrid method for biological applications, because it uses machine learning methods to infer a (mechanistic) dynamical system. However, using Breidenmoser and Wolkenhauer's distinction between phenomenological and mechanistic models, we would argue that, even though coated in terms like ``mechanistic", this method is nothing more than a phenomenological model: a curve is fitted to a derivative, instead of a time series. 

A similar critique can be applied to studies in which differential equations are learnt from model simulations \cite{patsatzis2023data}. For example, in one paper on hybrid models, Nardini et al. show how differential equations can be learned from agent-based simulations in order to predict how the latter type of model responds to parameter changes \cite{nardini2021learning}. One justification for this approach is that agent-based models might be computationally expensive to simulate. The approach also allows researchers to compare three approaches -- an agent based model, mean field equations derived from the agent based model, and a differential equation model learnt from the data provided by the agent based model--- in terms of the insights they offer and how well the methods approximate each other \cite{nardini2021learning}. In terms of the pluaralistic approach (which we advocate) the limitation of such a study is that it gives the (false) impression of generating three different views of a system, while it is primarily an exercise in deriving and estimating one model from another. In many studies describing hybrid approaches, data from real-world systems is not included. 

The danger here is that hybrid and equation-free modelling has a veneer of doing activities (1) through (3) but are, in fact, limited to activity (2). Real biological data is noisy, non-stationary and can be viewed in a multitude of ways. Understanding biology requires an openness to adopt different view points, rather than an attempt to close our approach down to one self-consistent framework. Fundamentally, any mathematical model or method should provide additional insights to a phenomenon, rather than cement a relationship between models. Hybrid approaches are not a substitute for, and are in can run counter to, an approach based onintegrative pluralism.

\section{How we can do better?}
\label{sec:dobetter}

The title of this paper is The Lost Art of Mathematical Modelling. In order to demonstrate why we think that something is lost, we have been critical of an emphasis of universalism and formalism in mathematical biology. It is in activity (2), we believe the subject can get lost. Our solution is to refocus on activity (1), creating new models. Smaldino takes a similar stand point but in social science \cite{smaldino2017models}. He argues that "it is time to focus on better practices for hypothesis generation. We need training programmes in model building and critique, plus consortia-building and funding programmes to invent and test measurements that make models tractable. Better methods will help us get the right answers; models and measurements will ensure we ask the right questions" \cite{smaldino2019better}. We fully agree with this position. We need to create more mathematical models in biology.

The way forward, we believe, is to view mathematical modelling in a more open way, one that admits biology is complex. This involves, as we saw in section \ref{sec:open}, creating lots of different models of a system in order to build up a broader understanding. There is a need, as Mitchell emphasises, for an approach based on integrative pluralism \cite{mitchell2006integration}, which emphasises  creating many different types of models (as is done in modelling fish locomotion; see figure 2).

\begin{figure}[t] \centering 
\includegraphics[scale=0.7]{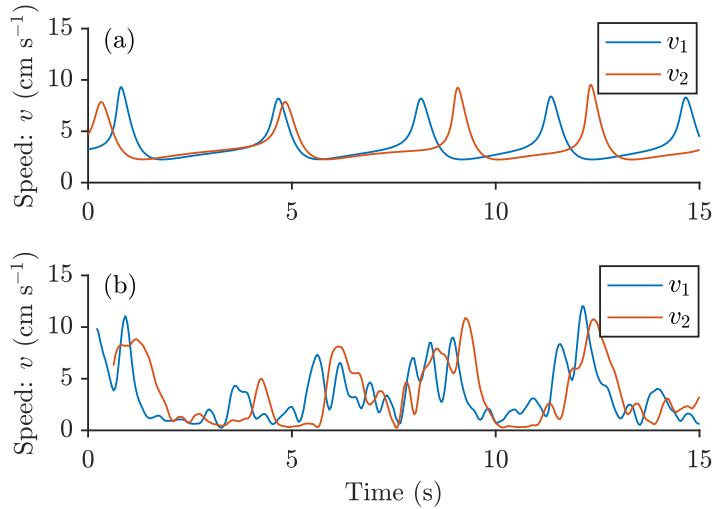}
\caption{Example of model compared with data from \cite{herbert2017predation}. Simulation of (a) a model of coupled differential equations describing the interaction of fish and (b) the speed of two fish over time for a period of 15 seconds of two guppies swimming in an arena.}
\label{fishmodel}
\end{figure}

Creating new models does not have to be a grandiose activity. For example, in a recent article we have developed a model of social burst and glide motion by combining a well-studied model of neuronal dynamics, the FitzHugh-Nagumo model, with a self-propelled model of fish motion \cite{gyllingberg2023using}. The FitzHugh-Nagumo equations model the membrane potential in a neuron, $V$, and the recovery variable, $W$ .\cite{fitzhugh1961impulses}. We found a way to couple this model to the velocity of the fish $v$ and a burst potential $b$, with the following equations:
\begin{align*} 
&\frac{db}{dt}=b-\frac{b^3}{3}-v+c   \\
&\frac{dv}{dt}=g(b)-kv,  
\end{align*}
where, $g(b)$ is given by 
\begin{equation*} \label{g-eq}
g(b)= a \Big(\frac{\arctan(z_1 (b-b_0))}{\pi}+\frac{1}{2}\Big). 
\end{equation*}
Figure \ref{fishmodel}(a) shows the burst and glide dynamics of two fish, interacting according to these dynamics, with their interactions coupled as a function of the distance between them. The model captures the responses of fish to each other, where one fish speeds up when the other fish moves ahead of it.


The challenge in this research was investigating the plausibility of a mechanism, i.e. whether the bursting of neurons, captured by the FitzHugh-Nagumo model, could be successfully related to burst and glide motion. At this stage of the research, comparison to data is less important \cite{bedau1999can}. Figure \ref{fishmodel}(b) shows that data collected from pairs of interacting guppies (as part of a study on effects of predation \cite{herbert2017predation}) is not entirely (or even very closely) described by the model, but there is something worth pursuing. Specifically, fish do respond to each other in their burst and glides and there is a need to relate this motion to neuronal mechanisms. The model we propose is certainly not the best way of predicting the time series in figure \ref{fishmodel}(b), but it could potentially be a way to identify a key mechanism in fish locomotion. 

Our point here is not to argue that this specific model is in itself a breakthrough, it is rather to give a feeling for what a refocussing on models might look like. It is OK to play around with different sets of equations and look to see, in a very loose way, whether they capture aspects of our understanding of a system.

\begin{figure}[t] \centering 
\includegraphics[scale=0.75]{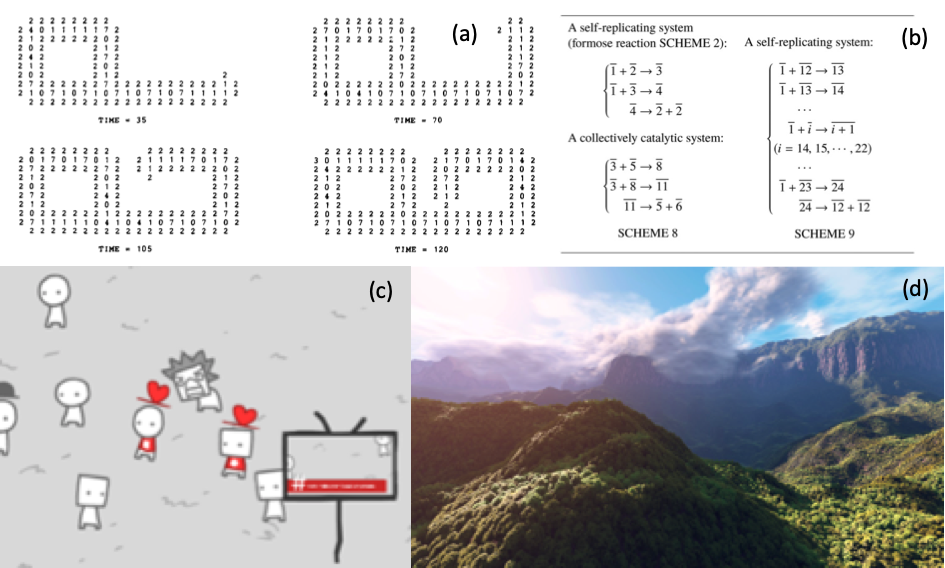}
\caption{Sketch examples of creating models of complex biological systems. (a) Langton’s (1984) loop is one of the first examples of a self-reproducing cellular automata \cite{langton1984self}; (b) Artificial chemistry gives insight in to both the origin of life and how complex components arise from simpler units \cite{liu2018mathematical}; (c) Nicky Case’s ‘We come what we behold’ is one example of an explorable game that allows the user to actively explore complexity \cite{ncase}; (d) Inigo Quilez work shows how realistic complex, fractal landscapes can be generated by a few lines of code \cite{quilez}.}
\label{newmodels}
\end{figure}

Creating new models can mean leaving our comfort zone. Some pointers in this direction include research in artificial life, developing online games where humans interact with simulations and investigating novel cellular automata (see figure \ref{newmodels} for examples). The common theme is an open-ended attempt to identify emergent phenomena, without ever trying to close the system with an exhaustive mathematical analysis. Instead of stifling theoretical development, we believe that mathematical biology should push to be more creative, to take risks and allow ourselves to be spectacularly wrong. 

%


It is the small insights, of how things fit together, which have constituted the biggest steps forward in mathematical biology. From Turing's work on morphogenesis \cite{turing1990chemical} and Hodgkin and Huxley's modelling of neuronal firing \cite{hodgkin1952quantitative}, through Hamilton's proposal of inclusive fitness \cite{hamilton1964genetical} and May's application of chaos theory in ecology \cite{may1974biological}, to Vicsek's model of collective motion \cite{vicsek1995novel} and Hogeweg's models of multi-level selection \cite{hogeweg2003multilevel}, it is the formulation of models (rather than their in-depth analysis) which has led to progress. Instead, of treating these existing models as sacred Platonic forms, which should be respected with deeper analysis, we should not be scared to look for new ideas and approaches. The inconvenient truth --- that biology is itself endlessly rich and varied and never subject to a final analysis --- is sometimes dismissed as an unrigorous approach. Such a situation is wrong. We need to rediscover the lost art of creating mathematical models.

\section{Epilogue}

This article is part of a special collection in memory of Edmund Crampin. Edmund was a close friend of one of the authors (David Sumpter) when they were PhD students. What David remembers most fondly about Edmund was his ability to be critical, both in a very deep way about his own work and constructively of the work of others. At that time, at the turn of the new millennium, Edmund was working on reaction-diffusion equations \cite{crampin1999reaction,crampin2002mode}, but was always torn as to whether the technical work he did truly contributed to biological insight. He felt that the experimental results collected at the time \cite{kondo1995reaction}, although supporting reaction-diffusion as a mechanism, did not entirely justify the extended theory he was working on. 

Rather than staying safely within the confines of one model, after his PhD, Edmund made sure he created new approaches to a variety of mathematical problems. When we look at his contributions --- ranging from multi-cellular modelling of the heart \cite{fink2011cardiac,smith2004multiscale}, through systems biology \cite{kohl2010systems} to biochemical reactions \cite{siekmann2012mcmc,crampin2004mathematical} and fundamentals of biophysics \cite{cudmore2021analysing,shahidi2022semantics}---we see an approach grounded in all three of the activities (modelling, anlaysis, comparison to data) we have discussed here. 

Edmund's self-insight has always stayed with David. And it is this spirit we have tried to adopt in this paper. Edmund may not have agreed with everything we have written --- we have taken a very strong position on what mathematical biology should be --- but he would have understood the need to be critical. Most of all, he would have enjoyed, over a long lunch or a vigorous walk, talking (and arguing) about the merits of different approaches to the subject he loved.


 \bibliographystyle{elsarticle-num} 
 \bibliography{cas-refs}





\end{document}